\documentclass[%
 reprint,
 superscriptaddress,
 amsmath,
 amssymb,
 amsfonts,
 aps,
 pre,
]{revtex4-2}

\usepackage{graphicx}
\usepackage{dcolumn}
\usepackage{bm}
\usepackage{mathtools}
\usepackage{upgreek}

\DeclareMathOperator\erfi{erfi}
\begin{document}

\title{Direct laser acceleration: A model for the electron injection \\ from the walls of a cylindrical guiding structure}

\author{P. Valenta}
\email{petr.valenta@eli-beams.eu}
\affiliation{ELI Beamlines Facility, Extreme Light Infrastructure ERIC, Za Radnicí 835, 252 41 Dolní Břežany, Czech Republic}

\author{D. Maslarova}
\affiliation{Institute of Plasma Physics, Czech Academy of Sciences, Za Slovankou 1782/3, 182 00 Praha 8, Czech Republic}
\affiliation{Faculty of Nuclear Sciences and Physical Engineering, Czech Technical University in Prague, Břehová 7, 115 19 Praha 1, Czech Republic}

\author{R. Babjak}
\affiliation{Institute of Plasma Physics, Czech Academy of Sciences, Za Slovankou 1782/3, 182 00 Praha 8, Czech Republic}
\affiliation{GoLP/Instituto de Plasma e Fusão Nuclear, Instituto Superior Técnico, Universidade de Lisboa, 1049-001 Lisbon, Portugal}

\author{B. Martinez}
\affiliation{GoLP/Instituto de Plasma e Fusão Nuclear, Instituto Superior Técnico, Universidade de Lisboa, 1049-001 Lisbon, Portugal}

\author{S. V. Bulanov}
\affiliation{ELI Beamlines Facility, Extreme Light Infrastructure ERIC, Za Radnicí 835, 252 41 Dolní Břežany, Czech Republic}
\affiliation{Kansai Photon Science Institute, National Institutes for Quantum Science and Technology, Umemidai 8-1-7, Kizugawa, Kyoto 619-0215, Japan}

\author{M. Vranić}
\affiliation{GoLP/Instituto de Plasma e Fusão Nuclear, Instituto Superior Técnico, Universidade de Lisboa, 1049-001 Lisbon, Portugal}

\date{\today}

\begin{abstract}

We use analytical methods and particle-in-cell simulation to investigate the origin of electrons accelerated by the process of direct laser acceleration driven by high-power laser pulses in preformed narrow cylindrical plasma channels. The simulation shows that the majority of accelerated electrons are originally located along the interface between the channel wall and the channel interior. The analytical model based on the electron hydrodynamics illustrates the underlying physical mechanism of the release of electrons from the channel wall when irradiated by an intense laser, the subsequent electron dynamics, and the corresponding evolution of the channel density profile. The quantitative predictions of the total charge of released electrons and the average electron density inside the channel are validated by comparison with the simulation results.
 
\end{abstract}

\maketitle

\section{\label{sec:introduction} Introduction}

Plasma-based electron accelerators have the potential to enable ground-breaking applications at significantly reduced size and cost in comparison with their radio-frequency counterparts. Apart from the well-established laser-wakefield accelerators \cite{tajima1979}, electrons can be also accelerated in plasmas by the so-called direct laser acceleration (DLA) \cite{pukhov1999a}.

The mechanism of DLA takes the advantage of a resonant process between the long-range “betatron-like” oscillations induced by self-generated radial electric and azimuthal magnetic fields in plasma and Doppler-shifted oscillations within a laser field. Numerous theoretical \cite{arefiev2012, arefiev2016, gong2020, jirka2020, wang2021, li2021a, yeh2021} as well as experimental \cite{gahn1999, willingale2013, singh2022, cohen2024} studies have already proven the feasibility and provided deeper understanding of this method. With petawatt-class power laser drivers, DLA has demonstrated the ability to generate high-energy electron beams with hundreds of $ \mathrm{nC} $ of charge \cite{gonoskov2014, ji2014, vranic2018, wang2019, hussein2021, shaw2021}. The envisioned applications of high-charge electron beams accelerated via the process of DLA include the production of bright gamma-ray and high-flux neutron sources \cite{rosmej2019, gunther2022}, exciting nuclear isomers \cite{feng2023}, seeding of the quantum electrodynamics (QED) cascades \cite{jirka2017}, and the guiding of positrons \cite{martinez2023, maslarova2023}.

As far as the positron guiding is concerned, previous works used the particle-in-cell (PIC) simulations to demonstrate the formation of dense electron beam by high-power lasers in plasma channels \cite{ji2014, vranic2018, wang2019, martinez2023}. This beam co-propagates with the laser and attracts positrons towards the channel center, enabling their acceleration to GeV energies. In this paper, we focus on the underlying physical mechanisms of the formation of such a guiding structure.

Our investigation employs both hydrodynamic analytical model and a full-scale PIC simulation. The analytical model illustrates the release of electrons from the channel wall when irradiated by the laser. It also captures the subsequent electron dynamics and the corresponding evolution of the channel density profile in both non-relativistic and relativistic limits. Furthermore, the analytical framework enables quantitative predictions of the number of electrons released from the channel wall and the corresponding average electron density inside the channel right after the passage of the laser pulse. 

We validate the aforementioned predictions by comparing them with the results obtained from the PIC simulation. The simulation showcases the generation of a high-energy (up to $ \approx 1.8 \ \mathrm{GeV} $) and high-charge ($ \approx 140 \ \mathrm{nC} $ with energy $ > 100 \ \mathrm{MeV} $) electron source through DLA driven by a kJ-class laser in a preformed narrow cylindrical plasma channel. It also demonstrates the formation of the guiding structure for positrons, which consists of electrons originating mostly at the interface between the interior of the channel and the channel wall. Further, we focus on determining the parameters of this guiding structure and its evolution in time. Our findings reveal that the plasma channel not only eliminates the diffraction spreading of the driving laser pulse but its wall also serves as a reservoir of electrons for DLA.

The remainder of this paper is organized as follows: In Sec.~\ref{sec:theory} we employ the analytical model based on the electron hydrodynamics illustrating the release of electrons from the channel wall. First, we introduce the initial set of equations for the analytical model (Sec.~\ref{sec:init_eq}). Subsequently, we explore the electron dynamics (Sec.~\ref{sec:el_dynamics}) and analyze the evolution of the channel density profile in both non-relativistic (Sec.~\ref{sec:nonrelativistic}) and relativistic (Sec.~\ref{sec:relativistic}) limits. Finally, we provide predictions for the number of electrons released from the channel wall by the laser pulse and the corresponding average electron density within the channel after the passage of the laser (Sec.~\ref{sec:laser}). In Sec.~\ref{sec:simulations} we present the setup and the results of the PIC simulation, including the comparison with the analytical predictions. Finally, we summarize the main results of this work in Sec.~\ref{sec:conclusion}.

\section{\label{sec:theory} Release of electrons from channel wall}

\subsection{\label{sec:init_eq} Initial considerations}

The interaction of an intense laser pulse with plasma targets critically depends on the ratio of the laser angular frequency, $ \omega_0 $, and the plasma frequency, $ \omega_p $. When interacting with overdense plasma (i.e., $ \omega_0 < \omega_p $), the absorption mechanisms of the laser pulse are determined by the plasma nonuniformity scale length, the pulse polarization, and the incidence angle (see Ref.~\cite{bulanov1994b} and the references cited therein). If the plasma nonuniformity is sufficiently large, the laser energy is absorbed mainly through the mechanism of electron ``vacuum heating'' \cite{brunel1987}, i.e., the flow of the plasma electrons breaks, their trajectories self-intersect, and their motion is of stochastic nature. Furthermore, a copious number of electrons initially located along the plasma boundary is released \cite{bulanov1994b}, which is of high relevance to both the electron injection for DLA and the formation of a guiding structure for positrons.

In this paper, we therefore limit our analysis to the laser propagating along the axis of a preformed plasma channel having a steep density gradient along its boundary. We also assume that the channel is cylindrically symmetric and sufficiently narrow, i.e., its diameter is comparable to the transverse size of the laser pulse. We use the equations of electron hydrodynamics in radial coordinate, $ r $, and time, $ t $, similarly as in Refs.~\citenum{dawson1959a, bulanov1994a, bulanov2004, vieira2012, bulanov2013a, vieira2014} (dynamics in planar geometry was studied also in Ref.~\citenum{valenta2023}). In dimensionless form, the equations can be written as
\begin{gather}
    \partial_t n_e + \frac{1}{r} \partial_r \left( r n_e v_e \right) = 0, \label{eq:1} \\
    \partial_t p_e + v_e \partial_r p_e = -E, \label{eq:2} \\
    \frac{1}{r} \partial_r \left( r E \right) = Z n_i - n_e. \label{eq:3}
\end{gather}
Here, $ r $ and $ t $ are normalized by $ c \omega_0^{-1} $ and $ \omega_0^{-1} $, respectively, where $ c $ is the velocity of light in vacuum. The radial components of velocity, $ v_e $, and momentum, $ p_e $, of the electron fluid component are measured in the units of $ c $ and $ m_e c $, respectively, where $ m_e $ is the electron mass. The electron velocity and momentum are related to each other by the expression $ v_e = p_e / (1 + p_e^2)^{1/2} $. The densities of electron, $ n_e $, and ion, $ n_i $, fluid components are normalized by the critical plasma density $ n_c = m_e \omega_0^2 / 4 \pi e^2 $, where $ e $ is the elementary charge. In the following calculations we assume that the ions are immobile (i.e., the ion density does not depend on time). The approximation of immobile ions is valid for early dynamics and should be revised for long laser pulses or long time evolution of the plasma. The radial component of the electric field, $ E $, is normalized by $ m_e \omega_0 c / e $. All quantities are assumed to be cylindrically symmetric.

By plugging $ n_e $ from Eq.~(\ref{eq:3}) into Eq.~(\ref{eq:1}) and integrating over $ r $ one gets
\begin{equation}\label{eq:4}
    \partial_t E + v_e \partial_r E = Z n_i v_e - \frac{v_e E}{r}.
\end{equation}
The system of Eqs.~(\ref{eq:1}), (\ref{eq:2}), and (\ref{eq:4}) can be solved, e.g., using Lagrange coordinates, $ r_0 $ and $ \tau $. The relation between the Euler and Lagrange coordinates is defined as $ r = r_0 + \rho $ and $ t = \tau $, where $ r_0 $ is the initial coordinate of the electron fluid element and $ \rho \left( r_0, \tau \right) $ is a displacement of the fluid element from $ r_0 $ during the time $ \tau $; thus $ \rho \left( r_0, 0 \right) = 0 $ and 
\begin{equation}\label{eq:5}
    \partial_{\tau} \rho = v_e = \frac{p_e}{\sqrt{1 + p_e^2}}. 
\end{equation}
For the partial derivatives with respect to the Euler coordinates we have
\begin{equation}\label{eq:6}
    \partial_r = \frac{1}{1 + \partial_{r_0} \rho} \partial_{r_0}, \quad \partial_t = \partial_{\tau} - \frac{v_e}{1 + \partial_{r_0} \rho} \partial_{r_0}. 
\end{equation}
In the Lagrange coordinates, Eqs.~(\ref{eq:1}), (\ref{eq:2}), and (\ref{eq:4}), respectively, thus become
\begin{gather}
    \partial_{\tau} n_e = -\frac{n_e}{1 + \partial_{r_0} \rho} \partial_{r_0 \tau} \rho - \frac{n_e}{r_0 + \rho} \partial_{\tau} \rho, \label{eq:7} \\
    \partial_{\tau} p_e = -E, \label{eq:8} \\
    \partial_{\tau} E = Z n_i \partial_{\tau} \rho - \frac{E}{r_0 + \rho} \partial_{\tau} \rho. \label{eq:9}
\end{gather}

We assume that the particle density inside the plasma channel is negligibly low and that the density gradient along the channel wall is sharp. Thus, we model its initial (i.e., at $ \tau = 0 $) transverse density profile with $ Z n_i \left( r_0 \right) = n_e \left( r_0, 0 \right) = n_0 \Theta \left( r_0 - R \right) $, where $ \Theta $ is the Heaviside step function [i.e., $ \Theta \left( x \right) = 1 $ for $ x > 0 $ and $ \Theta \left( x \right) = 0 $ for $ x \leq 0 $] and $ R $ represents the coordinate of the boundary between the interior of the channel and the channel wall. The solutions of Eqs.~(\ref{eq:7}) and (\ref{eq:9}), respectively, can be then written as 
\begin{gather}
    n_e = \frac{r_0}{r_0 + \rho} \frac{n_0 \Theta \left( r_0 - R \right)}{1 + \partial_{r_0} \rho}, \label{eq:10} \\
    E = \frac{n_0}{2} \Theta \left( r_0 + \rho - R \right) \frac{\left( r_0 + \rho \right)^2 - R^2}{r_0 + \rho} \nonumber \\ 
    - \frac{n_0}{2} \Theta \left( r_0 - R \right) \frac{r_0^2 - R^2}{r_0 + \rho}. \label{eq:11}
\end{gather}
One may see that the electron density, Eq.~(\ref{eq:10}), is inversely proportional to the Jacobian of the transformation from the Euler to Lagrange coordinates, $ J \left( r_0, \tau \right) = \left( r_0 + \rho \right) \left( 1 + \partial_{r_0} \rho \right) / r_0 $. When $ J $ vanishes (i.e., when $ \partial_{r_0} \rho \rightarrow -1 $), $ n_e $ tends to infinity. This indicates breaking of a plasma wave \cite{dawson1959a}.

\subsection{\label{sec:el_dynamics} Electron dynamics}

\begin{figure}[t]
\includegraphics[width=0.8\linewidth]{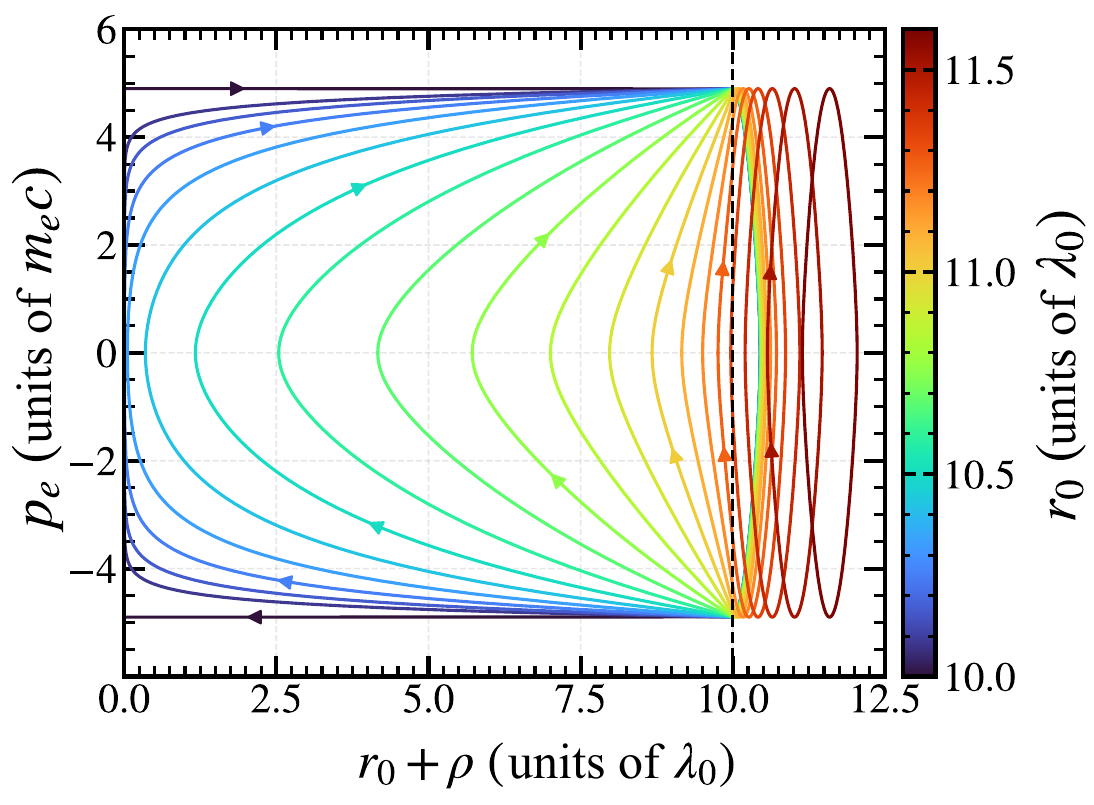}
\caption{(Color online). Contours of constant values of Hamilton function (\ref{eq:12}), where $ R = 10 \, \lambda_0 $ (black dashed line), $ n_0 = 1 \, n_c $, $ h = 5 $, and $ r_0 $ is varied from $ 10 $ to $ 11.6 \, \lambda_0 $. The arrows indicate the direction of motion of fluid elements and $ \lambda_0 = 2 \pi c / \omega_0 $ is the laser wavelength.}
\label{fig:1}
\end{figure}

Eqs.~(\ref{eq:5}) and (\ref{eq:8}) with the electric field given by Eq.~(\ref{eq:11}) can be cast into a Hamiltonian form with the Hamilton function
\begin{equation}\label{eq:12}
    \mathcal{H} = T + V,
\end{equation}
where $ T = \sqrt{1 + p_e^2} $ and
\begin{widetext}
\begin{equation}\label{eq:13}
    V = - \frac{n_0}{2} \Theta \left( r_0 + \rho - R \right) \left( R^2 \ln{\frac{r_0 + \rho}{R}} - \frac{\left( r_0 + \rho \right)^2 - R^2}{2} \right) - \frac{n_0}{2} \Theta \left( r_0 - R \right) \left( r_0^2 \ln{\frac{r_0 + \rho}{r_0}} - R^2 \ln{\frac{r_0 + \rho}{R}} + \frac{r_0^2 - R^2}{2} \right)
\end{equation}
\end{widetext}
are the kinetic and potential energies of electron fluid element, respectively. Since the Hamilton function (\ref{eq:12}) does not depend explicitly on time, the conservation of $ \mathcal{H} \left( \rho, p_e \right) = h \left( r_0 \right) $ gives a relationship between the electron momentum, $ p_e $, and the displacement, $ \rho $, along the trajectory determined by the value of $ h $,
\begin{equation}\label{eq:14}
    p_e = \pm \sqrt{\left( h - V \right)^2 - 1}.
\end{equation}
Here, the $ - $ and $ + $ signs in front of the square root correspond to the trajectories of electrons moving towards the channel axis and in the opposite direction, respectively. The phase portrait of the Hamiltonian system (\ref{eq:5}) and (\ref{eq:8}), which corresponds to the contours of constant values of Hamilton function (\ref{eq:12}), is shown in Fig.~\ref{fig:1}.

For given $ h $, one can identify three types of electron trajectories in the phase space depending on the value of $ r_0 $: (i) When $ r_0 \gg R $, the electrons perform oscillations around $ r_0 $ inside the channel wall with displacement $ \rho $ ranging from $ \rho^{-} $ (minimum value) to $ \rho^{+} $ (maximum value), where
\begin{equation}\label{eq:15}
    \rho^{\pm} \approx \pm \sqrt{\frac{2 \left( h - 1 \right)}{n_0}};
\end{equation}
(ii) when $ r_0 > R $ and $ r_0 + \rho^{-} < R $, the electrons within a single oscillation cycle cross the boundary between the channel wall and the channel interior, subsequently stop in the region inside the channel, and then return back to the channel wall. In such a case, the value of $ \rho^{-} $ is modified to
\begin{gather}
    \rho^{-} = -r_0 \nonumber \\ 
    + R \exp{\left( \frac{r_0^2}{r_0^2 - R^2} \ln{\frac{r_0}{R}} - \frac{2 \left( h - 1 \right)}{n_0 \left( r_0^2 - R^2 \right)} - \frac{1}{2} \right)}; \label{eq:16}
\end{gather}
and (iii), in the limiting case of $ r_0 = R $, $ \rho^{-} = -r_0 $, i.e., the electrons cross the boundary between the channel wall and the channel interior as well as the origin of the coordinate system and then return back to the channel wall.

\subsection{\label{sec:nonrelativistic} Non-relativistic limit}

\begin{figure}[t]
\includegraphics[width=0.8\linewidth]{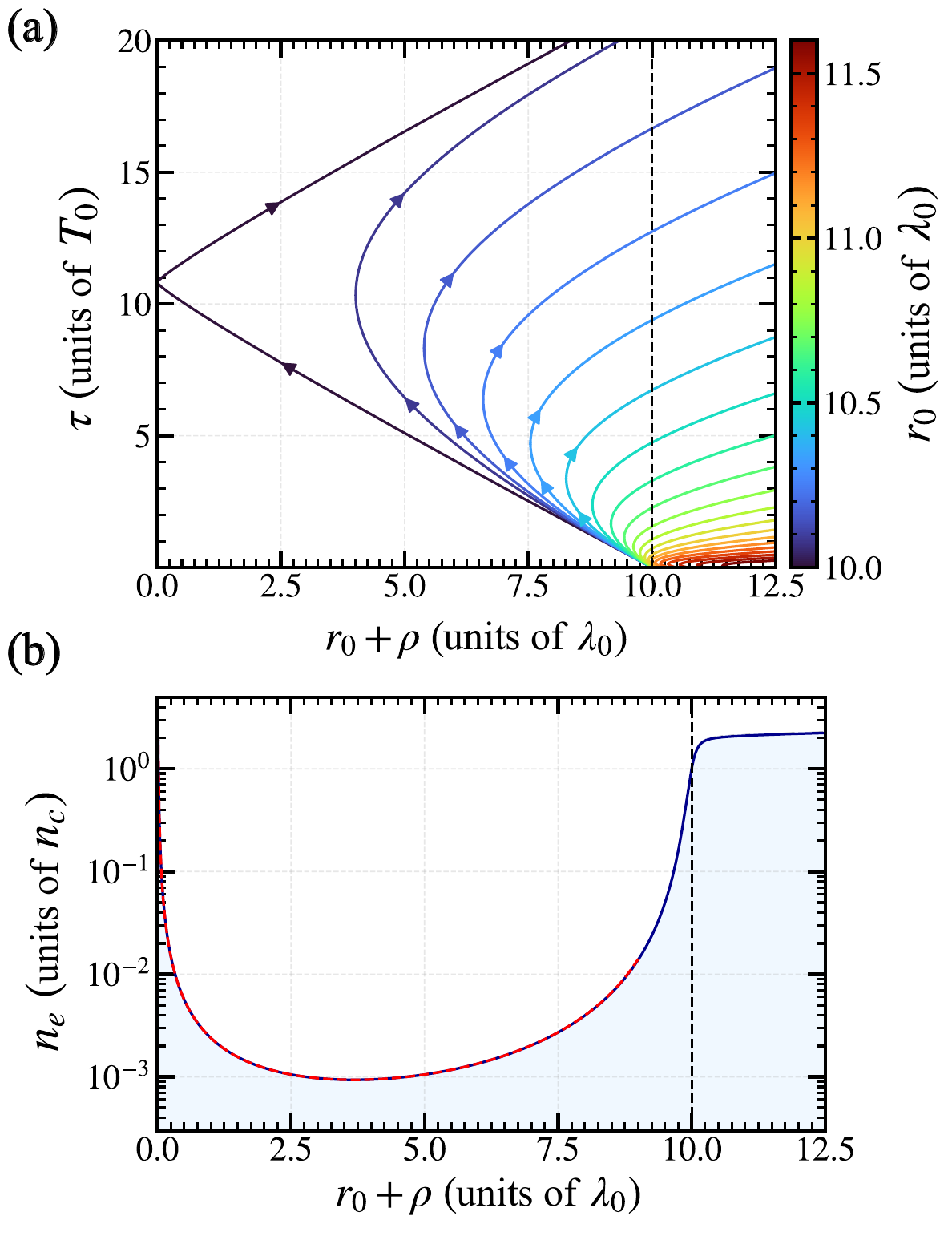}
\caption{(Color online). (a) Dependence of the displacement of electron fluid element, $ \rho $, on time, $ \tau $, in the non-relativistic limit according to Eq.~(\ref{eq:20}) with $ r_0 $ being varied from $ 10 $ to $ 11.6 \, \lambda_0 $. The arrows indicate the direction of motion of fluid elements and $ T_0 = 2 \pi / \omega_0 $ is the laser period. (b) Electron density, $ n_e $, according to Eq.~(\ref{eq:23}). The red dashed line corresponds to the approximation given by Eq.~(\ref{eq:24}). In both panels, $ R = 10 \, \lambda_0 $ (black dashed line), $ n_0 = 1 \, n_c $, and $ h = 1.5 $.}
\label{fig:2}
\end{figure}

In the non-relativistic limit (i.e., when $ p_e \approx v_e $), which is to a certain extent valid in the neighborhood of the point where the electron fluid element stops, $ r_0 + \rho^{-} $, Eq.~(\ref{eq:14}) can be written as
\begin{equation}\label{eq:17}
    \partial_{\tau} \rho = \pm \sqrt{2 \left( h - 1 \right) - 2 V}.
\end{equation}
Taking into account only the electron dynamics inside the channel (i.e., when $ r_0 > R $ and $ r_0 + \rho \leq R $), Eq.~(\ref{eq:13}) reduces to
\begin{equation}\label{eq:18}
    V = -\frac{n_0}{2} \left( r_0^2 \ln{\frac{r_0 + \rho}{r_0}} - R^2 \ln{\frac{r_0 + \rho}{R}} + \frac{r_0^2 - R^2}{2} \right)
\end{equation}
and Eq.~(\ref{eq:17}) therefore becomes
\begin{equation}\label{eq:19}
    \partial_{\tau} \rho = \pm \sqrt{n_0 \left( r_0^2 - R^2 \right) \ln{\frac{r_0 + \rho}{r_0 + \rho^{-}}}}.
\end{equation}
The solution of Eq.~(\ref{eq:19}) can be written in the following implicit form,
\begin{equation}\label{eq:20}
    \tau = \tau^{-} \pm \frac{\sqrt{\pi} \left( r_0 + \rho^{-} \right)}{\sqrt{n_0 \left( r_0^2 - R^2 \right)}} \erfi{\left( \sqrt{\ln{ \frac{r_0 + \rho}{r_0 + \rho^{-}} }} \right)},
\end{equation}
where
\begin{equation}\label{eq:21}
    \tau^{-} = \frac{\sqrt{\pi} \left( r_0 + \rho^{-} \right)}{\sqrt{n_0 \left( r_0^2 - R^2 \right)}} \erfi{\left( \sqrt{\ln{ \frac{r_0}{r_0 + \rho^{-}} }} \right)}
\end{equation}
is the time it takes for the electron fluid element to travel from $ r_0 $ to $ r_0 + \rho^{-} $ and $ \erfi{\left( x \right)} $ is the imaginary error function. The dependence of $ \rho $ on $ \tau $ according to Eq.~(\ref{eq:20}) is depicted in panel (a) of Fig.~\ref{fig:2}, where one can clearly recognize the electron trajectories of types (ii) and (iii) mentioned in Sec.~\ref{sec:el_dynamics}.

Using the asymptotic expansion of $ \erfi{\left( x \right)} $ for $ \rho \rightarrow \rho^{-} $ one obtains
\begin{equation}\label{eq:22}
    \rho = -r_0 + \left( r_0 + \rho^{-} \right) \exp{\left[ \frac{n_0 \left( r_0^2 - R^2 \right) }{4} \left( \frac{\tau - \tau^{-}}{r_0 + \rho^{-}} \right)^2 \right]}.
\end{equation}
By plugging the derivative of $ \rho $ [given by Eq.~(\ref{eq:22})] with respect to $ r_0 $ into Eq.~(\ref{eq:10}), one obtains the evolution of electron density inside the channel. Its expression for $ \tau \rightarrow \tau^{-} $ takes the following simple form,
\begin{equation}\label{eq:23}
    n_e = \frac{n_0 \left( R^2 - r_0^2 \right) \Theta \left( r_0 - R \right)}{2 \left( r_0 + \rho^{-} \right)^2 \ln{\left[ \left( r_0 + \rho^{-} \right) / r_0 \right]}}.
\end{equation}
For $ r_0 + \rho^{-} \rightarrow 0 $ (i.e., near the channel axis), the electron density given by Eq.~(\ref{eq:23}) can be approximated (and transformed back to Euler coordinate $ r $) as
\begin{equation}\label{eq:24}
    n_e \approx \frac{h - 1}{\left[ r \ln{\left( r / R \right)} \right]^2}.
\end{equation}
One may see that for $ r \rightarrow 0 $ the electron density tends to infinity. However, the singularity is integrable, i.e., the total number of electrons near the axis remains finite. The electron density profile described by Eq.~(\ref{eq:23}) as well as the approximation of the electron density filament formed along the channel axis given by Eq.~(\ref{eq:24}) are shown in panel (b) of Fig.~\ref{fig:2}.

\subsection{\label{sec:relativistic} Relativistic limit}

We now shift our focus to the relativistic case relevant to DLA, where relativistic laser intensities inherently produce high-energy electrons. In the relativistic limit, the dependence of the displacement on time for each electron fluid element can be obtained by integration of Eq.~(\ref{eq:5}) along the phase space trajectory of the fluid element. Parametrizing a single oscillation cycle as
\begin{equation}\label{eq:25}
    \rho \left( r_0, s \right) = \left[ \rho^{-} \Theta \left( \sin{s} \right) - \rho^{+} \Theta \left( -\sin{s} \right) \right] \sin{s}
\end{equation}
with $ s \in \left[ 0, \, 2 \pi \right] $, one gets the relationship in the following implicit form,
\begin{gather}
    \tau = \oint_{0}^{s} \frac{h - V \left( \rho \left( r_0, s^{\prime} \right) \right)}{\sqrt{\left[h - V \left( \rho \left( r_0, s^{\prime} \right) \right) \right]^2 - 1}} \nonumber \\
    \times \left[ \rho^{+} \Theta \left( -\sin{s^{\prime}} \right) - \rho^{-} \Theta \left( \sin{s^{\prime}} \right) \right] \left| \cos{s^{\prime}} \right| \, \mathrm{d} s^{\prime}. \label{eq:26}
\end{gather}
Here, the electron potential energy, $ V $, is given by Eq.~(\ref{eq:13}).

\begin{figure}[t]
\includegraphics[width=0.875\linewidth]{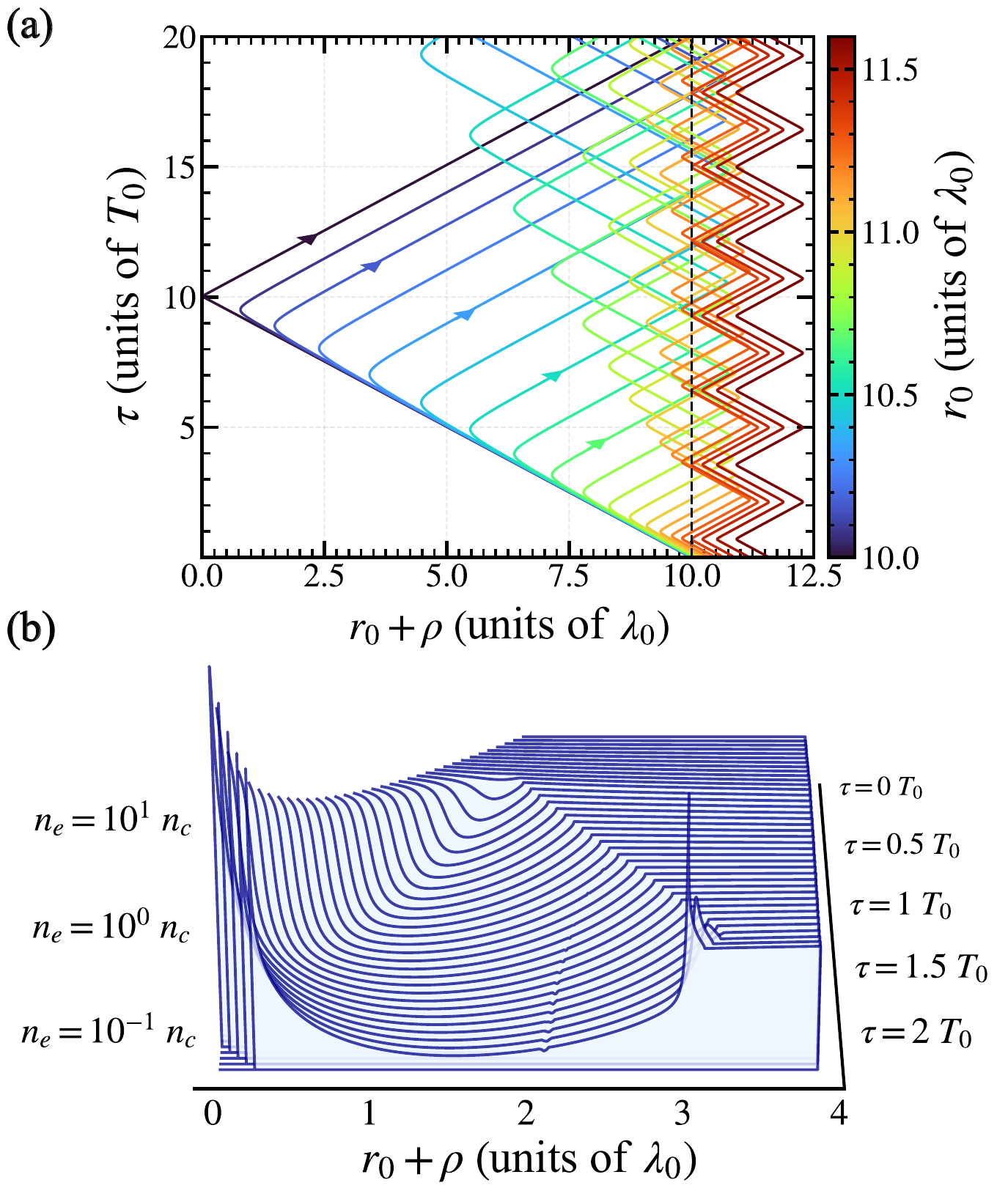}
\caption{(Color online). (a) Dependence of the displacement of electron fluid element, $ \rho $, on time, $ \tau $, in the relativistic limit according to Eq.~(\ref{eq:26}) with $ R = 10 \, \lambda_0 $ (black dashed line) and $ r_0 $ being varied from $ 10 $ to $ 11.6 \, \lambda_0 $. The arrows indicate the direction of motion of fluid elements. (b) Electron density, $ n_e $, evolution in time, $ \tau $, obtained from Eqs.~(\ref{eq:10}) and (\ref{eq:26}) with $ R = 2 \, \lambda_0 $ in order to show the density profile along the channel axis before the wave breaking occurs. In both panels, $ n_0 = 1 \, n_c $ and $ h = 10 $.}
\label{fig:3}
\end{figure}

Panel (a) of Fig.~\ref{fig:3} shows the dependence of $ \rho $ on $ \tau $ according to Eq.~(\ref{eq:26}) with the integral on the right-hand side being evaluated numerically. Here, one can recognize all three types of electron trajectories mentioned in Sec.~\ref{sec:el_dynamics}. Contrary to the non-relativistic limit presented in Sec.~\ref{sec:nonrelativistic}, one may see that the dependence has characteristic ``sawtooth-like'' form because the velocity of the electron fluid elements during subsequent half-periods is close to $ \pm c $. 

After one half-period, when the individual elements return back to the channel wall, they begin to intersect each other's paths, which indicates the presence of a multi-stream flow and, possibly, the electron density singularities (e.g., cusps, folds, etc.). These orbit self-intersections arise from the nonuniformity of ion density distribution at the interface between the channel interior and its wall. We note that beyond this point the hydrodynamic description of Eqs.~(\ref{eq:1})--(\ref{eq:3}) becomes inadequate.

The evolution of electron density in $ \tau $, which can be obtained from Eqs.~(\ref{eq:10}) and (\ref{eq:26}), is illustrated in panel (b) of Fig.~\ref{fig:3}. Since this approach is taking into account also the dynamics inside the channel wall, apart from the shock wave propagating towards the channel axis, one may see also a rarefaction wave propagating towards the channel wall. Prior to the occurrence of a multi-stream flow, one can observe the formation of thin and dense electron filament on the channel axis (similarly as in the non-relativistic case) as well as the onset of the plasma wave breaking on the interface between the channel interior and the channel wall.

\subsection{\label{sec:laser} Laser interaction with channel wall}

When interacting with laser pulse, the initial energy of an electron located in the vicinity of $ R $ can be approximated as $ h = \sqrt{1 + a_{\mathrm{eff}}^2} $, where $ a_{\mathrm{eff}} $ stands for the effective value of the normalized pulse amplitude characterizing the strength of the interaction, i.e., $ h \approx 1 + a_{\mathrm{eff}}^2 / 2 $ when $ a_{\mathrm{eff}} \ll 1 $ (non-relativistic limit) and $ h \approx a_{\mathrm{eff}} $ when $ a_{\mathrm{eff}} \gg 1 $ (relativistic limit).

According to Eq.~(\ref{eq:15}), an electron is released from the channel wall to the interior of the channel when $ \rho^{-} < R - r_0 $. Therefore, the maximum number of electrons, $ N_e $, being released from the channel wall due to the action of the laser pulse at a given longitudinal coordinate $ x $, is
\begin{equation}\label{eq:27}
    \mathrm{d}_x N_e \approx \pi \kappa \left( 2 R \sqrt{n_0} + \kappa \right) \left( \frac{c}{\omega_0} \right)^2 n_c
\end{equation}
with
\begin{equation}\label{eq:28}
    \kappa = \begin{dcases}
    a_{\mathrm{eff}} & \text{if } a_{\mathrm{eff}} \ll 1 \\
    \sqrt{2 \left( a_{\mathrm{eff}} - 1 \right)} & \text{if } a_{\mathrm{eff}} \gg 1
    \end{dcases}.
\end{equation}

If the laser pulse is sufficiently long, a substantial portion of these electrons are rotated along the channel axis by the magnetic component of the pulse, making them available for injection \cite{jiang2018} and subsequent acceleration in the longitudinal direction by DLA. The rest eventually returns back to the channel wall. Neglecting the electrons trapped by the laser field, the average electron number density inside the channel right after the passage of the laser is thus given by
\begin{equation}\label{eq:29}
    \langle n_e \rangle \approx \frac{\kappa}{R^2} \left( 2 R \sqrt{n_0} + \kappa \right).
\end{equation}
The value of $ a_{\mathrm{eff}} $ is studied using the PIC simulation in the following section.

\section{\label{sec:simulations} Particle-in-cell simulation}

\subsection{\label{sec:setup} Simulation setup}

The parameters of the PIC simulation are defined as follows: The driving laser pulse is characterized by the central wavelength $ \lambda_0 = 2 \pi c / \omega_0 $ and dimensionless amplitude $ a_0 = e E_0 / m_e \omega_0 c = 50 $, where $ E_0 $ is the peak amplitude of the laser electric field in vacuum. In this case we may neglect the effects of radiation reaction force on injected electrons \cite{bulanov2011a, jirka2020}. The pulse has Gaussian profile in the radial direction with the beam waist $ w_0 = 10 \, \lambda_0 $ and a Gaussian-like $ 5^{\mathrm{th}} $ order symmetric polynomial profile in the longitudinal direction with duration $ \tau_0 = 50 \, T_0 $ (in the full-width at half-maximum of the laser electric field profile), where $ T_0 = \lambda_0 / c $ is the cycle period of the laser. In what follows we assume $ \lambda_0 = 1 \, \mathrm{\upmu m} $, i.e., the total energy, peak power, and peak intensity of the laser pulse in vacuum are, respectively, $ \mathcal{E}_0 \approx 0.7 \, \mathrm{kJ} $, $ P_0 \approx 5.4 \, \mathrm{PW} $, and $ I_0 \approx 3.4 \times 10^{21} \, \mathrm{W / cm^2} $. 

\begin{figure}[t]
\includegraphics[width=0.95\linewidth]{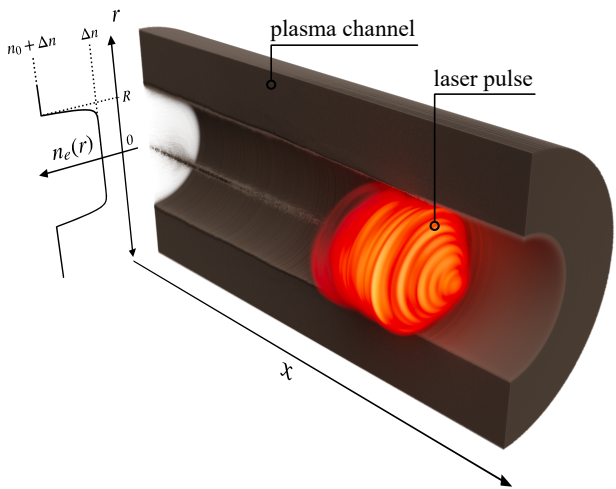}
\caption{(Color online). Setup of the PIC simulation: The laser pulse (orange) propagates in preformed plasma channel (gray) with the initial radial density profile, $ n_e \left( r \right) $, defined by Eq.~\ref{eq:30}.}
\label{fig:4}
\end{figure}

The laser pulse propagates in a preformed narrow cylindrically symmetric channel, which consists of a fully ionized plasma. Its initial electron density profile dependence on the radial coordinate is given by
\begin{equation}\label{eq:30}
    n_e(r) = \begin{dcases}
    n_0 \exp{\left( \frac{r - R}{\lambda_0 / 2} \right)} + \Delta n & \text{if } r \leq R \\
    n_0 + \Delta n & \text{if } r > R
    \end{dcases},
\end{equation}
where $ R = w_0 $ is the channel radius, $ n_0 = 10 \, n_c $, and $ \Delta n = 10^{-3} \, n_c $ (see Fig.~\ref{fig:4}). In order to account for the effect of laser pre-pulse, the plasma along the interface between the channel wall and the channel interior is slightly pre-expanded; the characteristic length of the exponential profile is chosen such that $ n_e = n_c $ at $ r \approx R - \lambda_0 $. In addition, there is a conical opening at the front side of the channel, which facilitates the transmission of the laser pulse into the channel. The initial ion density profile is chosen such that the condition of plasma quasi-neutrality is fulfilled. The laser pulse propagates along the channel axis (denoted as the $ x $-axis) and is linearly polarized along the $ y $-axis. Its focal plane is located at the entrance to the plasma channel.

The PIC simulation is carried out in a cylindrical geometry with the azimuthal Fourier decomposition of the electromagnetic field components (sometimes referred to as quasi-3D geometry) using the OSIRIS framework \cite{fonseca2002}. This simulation approach is suitable for systems close to the cylindrical symmetry, where a low number of modes is sufficient. While the quasi-particles are allowed to move in the three-dimensional Cartesian geometry, the electromagnetic fields are calculated only on two-dimensional lattices \cite{lifschitz2009, davidson2015}. In the simulation, we decompose the electromagnetic fields into two modes; the first mode account for the axisymmetric self-generated channel fields, and the second mode for the non-axisymmetric linearly polarized field of the laser. 

The simulation utilizes moving window technique; the window, which moves at the velocity of $ c $, has dimensions of $ 250 $ and $ 25 \, \lambda_0 $ in the longitudinal and radial directions, respectively. The underlying Cartesian grid is uniform with the resolution of $ 40 $ cells per $ \lambda_0 $ in both directions. The simulation is evolved over the time interval of $ 2 \times 10^{3} \, T_0 $. The plasma is cold and collisionless, represented with electron and ion quasi-particles. Initially, there are $ 32 $ quasi-particles per grid cell for each species. The electromagnetic field evolution is calculated using a high-order finite-difference Maxwell solver \cite{li2017a}, whereas the equations of motion for quasi-particles are solved using the Boris algorithm \cite{boris1971}. Absorbing boundary conditions are applied on each of the simulation window boundaries for both the electromagnetic fields and quasi-particles.

\subsection{\label{sec:results} Simulation results}

\begin{figure}[t!]
    \centering
    \includegraphics[width=0.95\linewidth]{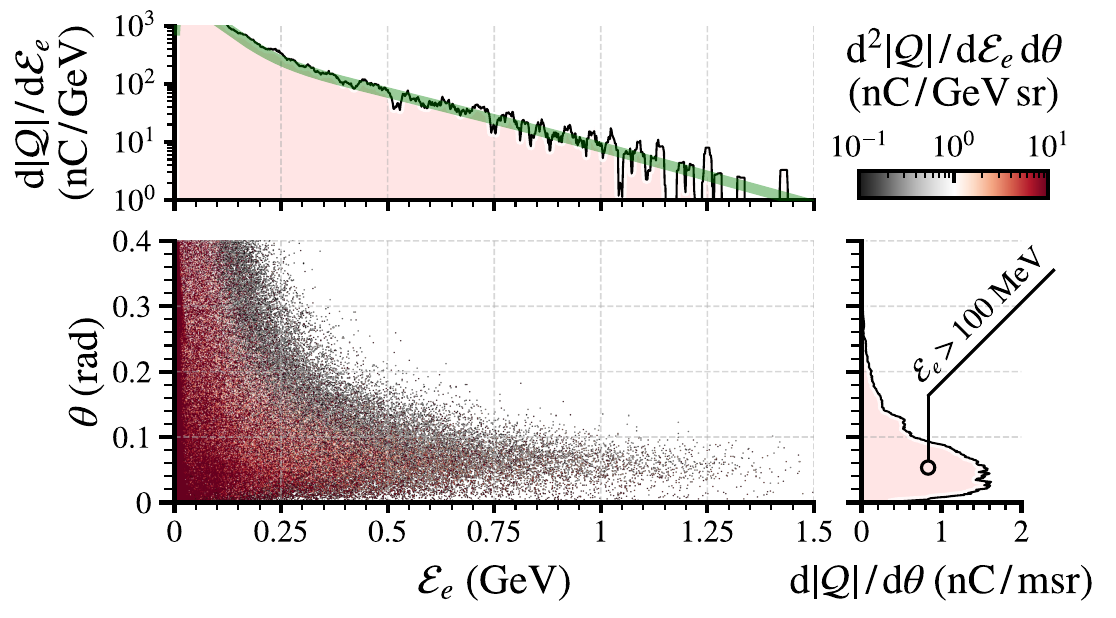}
    \caption{(Color online). (top) Electron energy spectrum, $ \mathrm{d} \left| \mathcal{Q} \right| / \mathrm{d} \mathcal{E}_e $, (bottom left) charge density of electrons with respect to their kinetic energy and propagation angle, $ \mathrm{d}^2 \left| \mathcal{Q} \right| / \mathrm{d} \mathcal{E}_e \mathrm{d} \theta $, and (bottom right) integrated charge density of electrons with kinetic energy $ > 100 \, \mathrm{MeV} $ with respect to their propagation angle, $ \mathrm{d} \left| \mathcal{Q} \right| / \mathrm{d} \theta $, at the end of the PIC simulation. All electrons inside the simulation box are taken into account. The solid green line in the upper panel shows a two-temperature Maxwell-Boltzmann distribution approximating the electron energy spectrum with temperatures equal to $ 50 $ and $ 200 \, \mathrm{MeV} $.}
    \label{fig:5}
\end{figure}

The properties of electrons inside the simulation box at the end of the PIC simulation (i.e., at $ t = 2 \times 10^3 \, T_0 $) are shown in Fig.~\ref{fig:5}. The electrons have thermal energy spectrum, which can be approximated with a two-temperature Maxwell-Boltzmann distribution with temperatures equal to $ 50 \, \mathrm{MeV} $ for low-energy electrons and $ 200 \, \mathrm{MeV} $ for the high-energy tail. The cut-off energy of electrons increases linearly with the laser propagation distance and saturates at the level of $ \approx 1.8 \, \mathrm{GeV} $. According to DLA scaling laws for this regime \cite{babjak2024}, the cut-off energy of electrons is determined by the combination of the electron's transverse oscillation amplitude, $ r_{\mathrm{max}} $, and the ambient electron density, i.e., $ \mathcal{E}_{\mathrm{max}} / m_e c^2 = 2 \mathcal{I}^2 n_c / n_e $ with $ \mathcal{I} = 1 + \pi^2 r_{\mathrm{max}}^2 n_e / \lambda_0^2 n_c $. The value obtained from the simulation can be recovered with $ n_e = \Delta n $ and $ r_{\mathrm{max}} = 6 \, \lambda_0 $ (which is lower than $ R $ because of the pre-expanded channel wall).

The average angle between the channel axis and the propagation direction of the electrons from the high-energy tail is $ \approx 50 \, \mathrm{mrad} $, whereas the lower-divergence electrons are far less present. This is a characteristic feature of DLA, which manifests itself through forked structures in electron spectra \cite{shaw2017, shaw2018, king2021}. The divergence of high-energy electrons corresponds well with $ \theta \approx \arctan \sqrt{2 m_e c^2 \mathcal{I} / \mathcal{E}_e} $, where $ \mathcal{E}_e $ denotes the electron energy \cite{jirka2020}.

\begin{figure}[t!]
    \centering
    \includegraphics[width=0.95\linewidth]{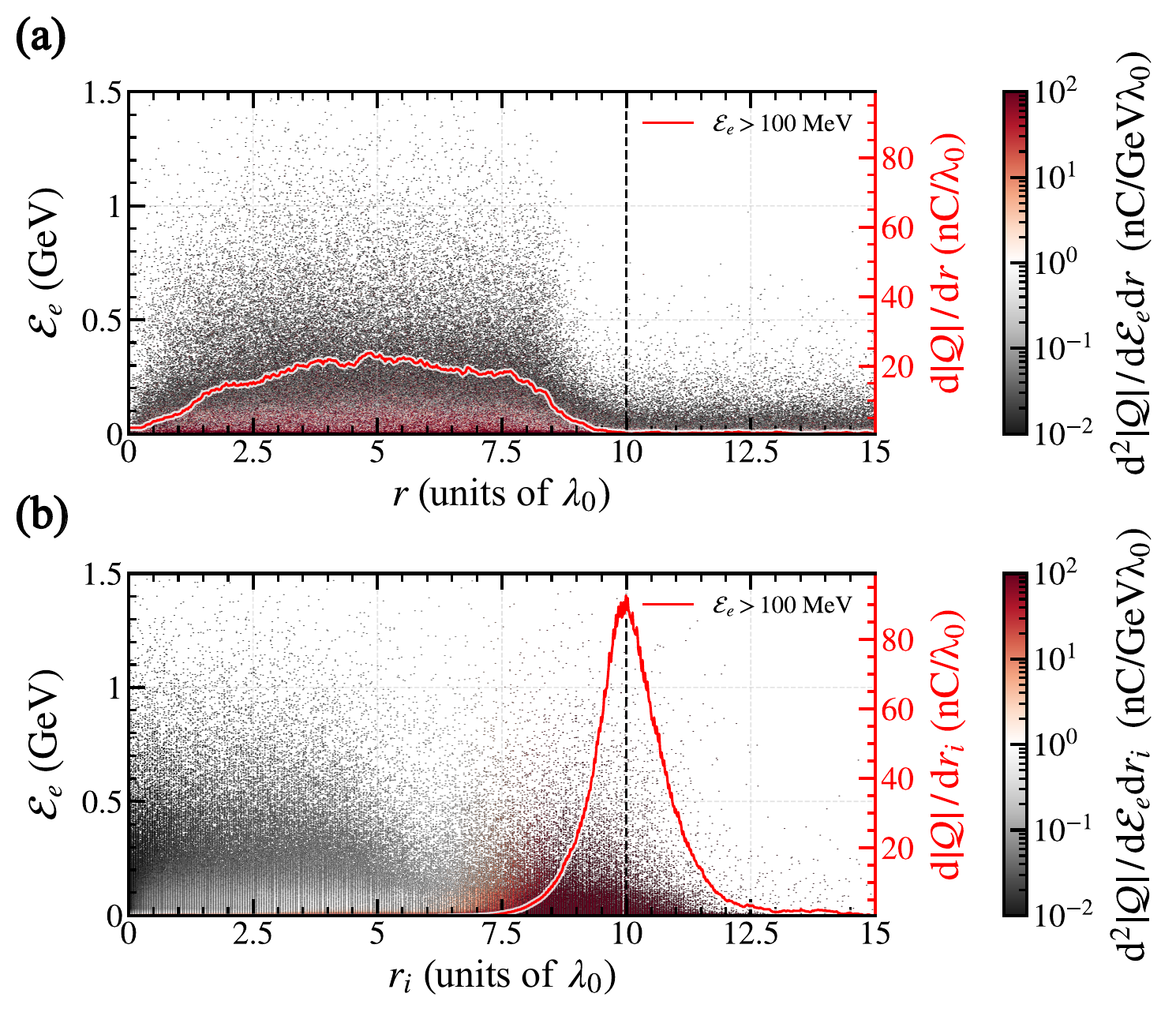}
    \caption{(Color online). (a) Distribution of electron charge with respect to their kinetic energy and radial coordinate, $ \mathrm{d}^2 \left| \mathcal{Q} \right| / \mathrm{d} \mathcal{E}_e \mathrm{d} r $, at the end of the PIC simulation. All electrons inside the simulation box are taken into account. The solid red line shows the distribution of charge of high-energy ($ \mathcal{E}_e > 100 \, \mathrm{MeV} $) electrons with respect to their radial coordinate, $ \mathrm{d} \left| \mathcal{Q} \right| / \mathrm{d} r $. (b) Same as (a) with the radial coordinate of electrons, $ r $, being replaced by their initial radial coordinate, $ r_i $. The black dashed lines in both panels mark the channel radius, $ R $.}
    \label{fig:6}
\end{figure}

The distribution of the electron charge with respect to their kinetic energy, $ \mathcal{E}_e $, and radial coordinate, $ r $, at the end of the simulation is depicted in panel (a) of Fig.~\ref{fig:6}. The panel shows as well the distribution of the charge of high-energy ($ \mathcal{E}_e > 100 \, \mathrm{MeV} $) electrons with respect to $ r $. It can be observed that the electrons from both low-energy and high-energy spectrum parts are distributed across the entire area of the plasma channel, contributing to the beam loading effect. Most high-energy electrons ($ \approx 25 \, \mathrm{nC} / \lambda_0 $) are located around the radial coordinate $ r \approx 5 \, \lambda_0 $, while fewer of them are found along the channel axis and the channel wall. The total charge of electrons with $ \mathcal{E}_e > 100 \, \mathrm{MeV} $ inside the channel is $ \approx 140 \, \mathrm{nC} $.

Panel (b) of Fig.~\ref{fig:6} displays the same distribution as panel (a) with the radial coordinate of electrons being replaced by their initial radial coordinate, $ r_i $. One can see that the majority of electrons accelerated to $ > 100 \, \mathrm{MeV} $ originate in the vicinity of the interface between the channel interior and the channel wall, with the peak charge density reaching up to $ \approx 90 \, \mathrm{nC} / \lambda_0 $; the initial radial coordinate of $ \approx 75 \, \% $ of these electrons falls within the range $ R - \lambda_0 $ to $ R + \lambda_0 $. This numerical result (i.e., that most of the accelerated electrons come from the wall of the plasma channel) is in line with the main assumption of the analytical model presented in Sec.~\ref{sec:theory}.

\begin{figure}[t]
    \centering
    \includegraphics[width=0.95\linewidth]{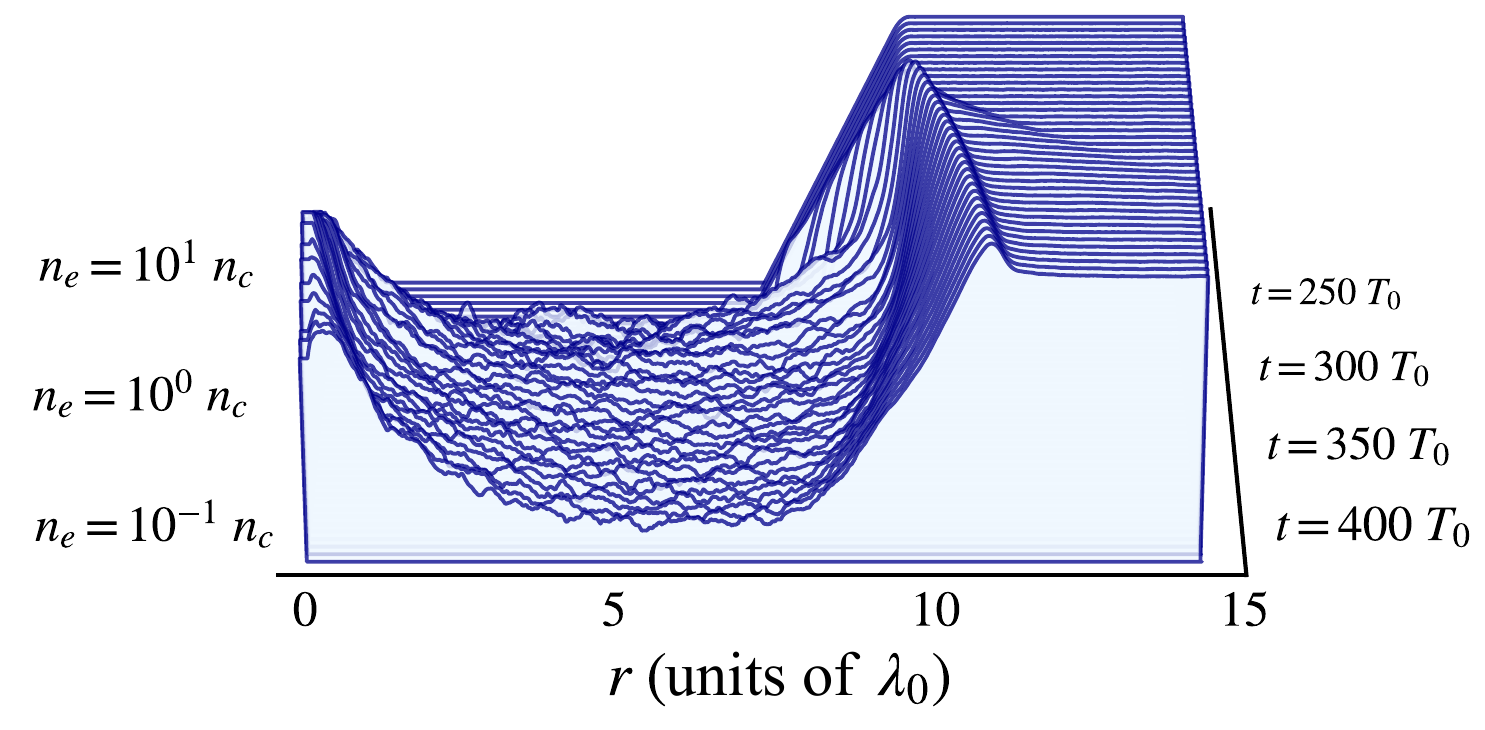}
    \caption{(Color online). Evolution of electron density, $ n_e $, due to the interaction with the laser pulse in radial coordinate, $ r $, and time, $ t $, at the longitudinal coordinate $ x = 200 \ \lambda_0 $ obtained from the PIC simulation.}
    \label{fig:7}
\end{figure}

Figure~\ref{fig:7} shows the evolution of the electron density due to the interaction with the laser pulse in radial coordinate and time at the longitudinal coordinate $ x = 200 \, \lambda_0 $ obtained from the PIC simulation. One can recognize the main features captured by the analytical model; the release of electrons from the channel wall, the formation of a high-density filament along the channel axis, and the plasma wave breaking along the interface between the channel interior and the channel wall. In addition to the model, the evolution of electron density obtained from the simulation reveals a slow expansion of the channel radius due to the ion motion.

\begin{figure}[t]
    \centering
    \includegraphics[width=0.85\linewidth]{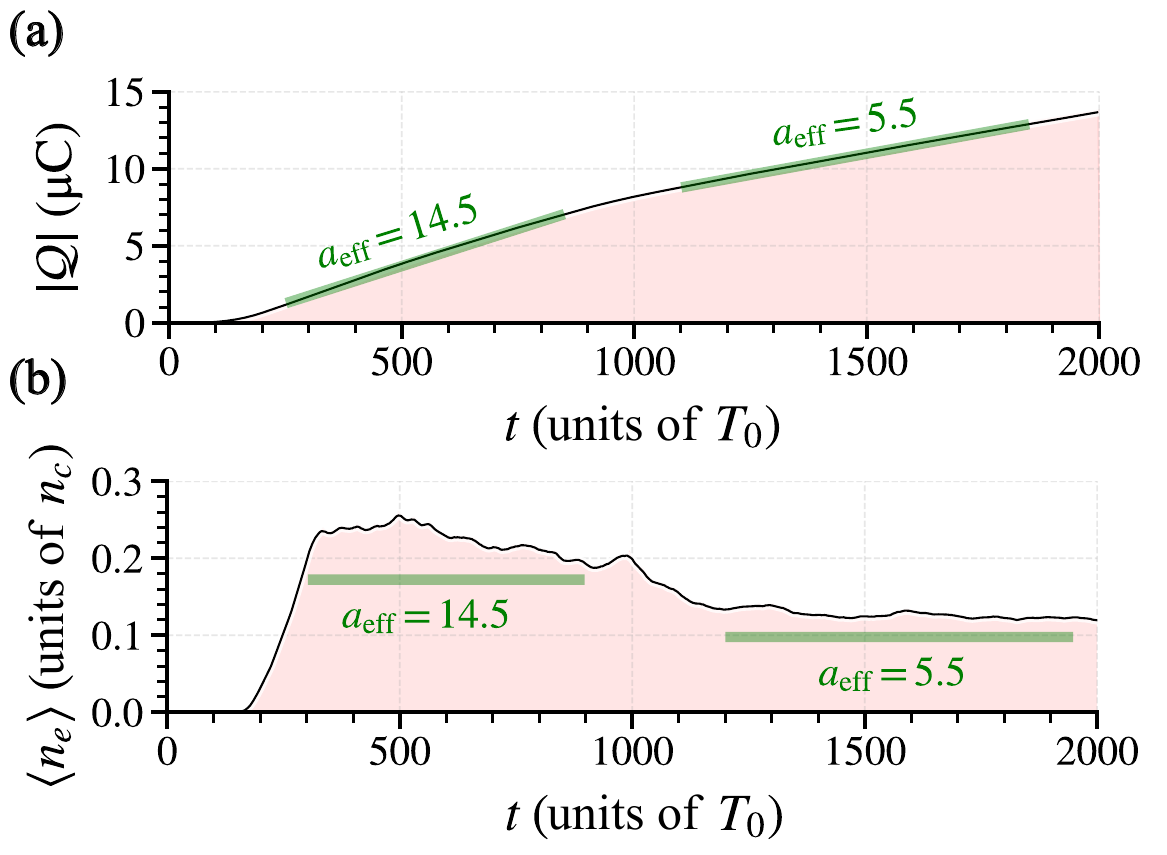}
    \caption{(Color online). (a) Evolution of the total charge, $ \mathcal{Q} $, of the electrons released from the channel wall in time, $ t $, obtained from the PIC simulation. The solid green lines show the approximation given by Eq.~(\ref{eq:27}) with corresponding values of $ a_{\mathrm{eff}} $. (b) Evolution of the average electron number density inside the channel, $ \langle n_e \rangle $, right after the passage of the laser pulse in time obtained from the PIC simulation. The solid green lines show the approximation given by Eq.~(\ref{eq:29}) with the same values of $ a_{\mathrm{eff}} $ as in (a).}
    \label{fig:8}
\end{figure}

Let us now consider only the electrons initially positioned in the channel wall (the boundary between the interior and the wall of the channel shifts due to the pre-expanded channel wall, i.e., $ r_i \geq R - \lambda_0 $) and later released to the interior of the channel (i.e., $ r < R - \lambda_0 $). The time evolution of the total charge of these electrons is displayed in panel (a) of Fig.~\ref{fig:8}. As can be seen, the rate at which the electrons are released is highest soon after the pulse enters the channel ($ \approx 5 \, \mathrm{nC} / T_0 $) and then continuously decreases (down to $ \approx 2.5 \, \mathrm{nC} / T_0 $) over the course of the time span captured by the PIC simulation due to the energy loss of the driving laser. In total, $ \approx 15 \, \mathrm{\upmu C} $ of the electron charge is released until the end of the simulation. 

The value of $ a_{\mathrm{eff}} $, discussed in Sec.~\ref{sec:laser}, can be found by fitting the evolution of the rate at which the electrons are released from the channel wall [shown in panel (a) of Fig.~\ref{fig:8})] by the formula of Eq.~(\ref{eq:27}). As can be seen, the rate can be approximated with $ a_{\mathrm{eff}} = 14.5 $ and $ 5.5 $ for the first and second halves of the simulation, respectively. These values are comparable to the normalized amplitude of the laser pulse at the interface between the channel wall and the channel interior at the corresponding time instants. 

Panel (b) of Fig.~\ref{fig:8} shows the time evolution of the average electron number density inside the plasma channel right after the passage of the laser pulse. We can see that the average density is about twice as high in the first half of the simulation (maximum $ n_e \approx 0.25 \, n_c $) as in the second half (maximum $ n_e \approx 0.13 \, n_c $), which is in agreement with the decrease of the rate at which the electrons are released from the channel wall. The average density can be approximated using the formula of Eq.~(\ref{eq:29}) with the same values of $ a_{\mathrm{eff}} $ as before; one obtains $ \langle n_e \rangle \approx 0.17 $ for $ a_{\mathrm{eff}} = 14.5 $ and $ \langle n_e \rangle \approx 0.1 $ for $ a_{\mathrm{eff}} = 5.5 $. The values obtained analytically are slightly lower because the analytical model does not take into account the pre-expanded channel wall.

\begin{figure}[t]
    \centering
    \includegraphics[width=0.9\linewidth]{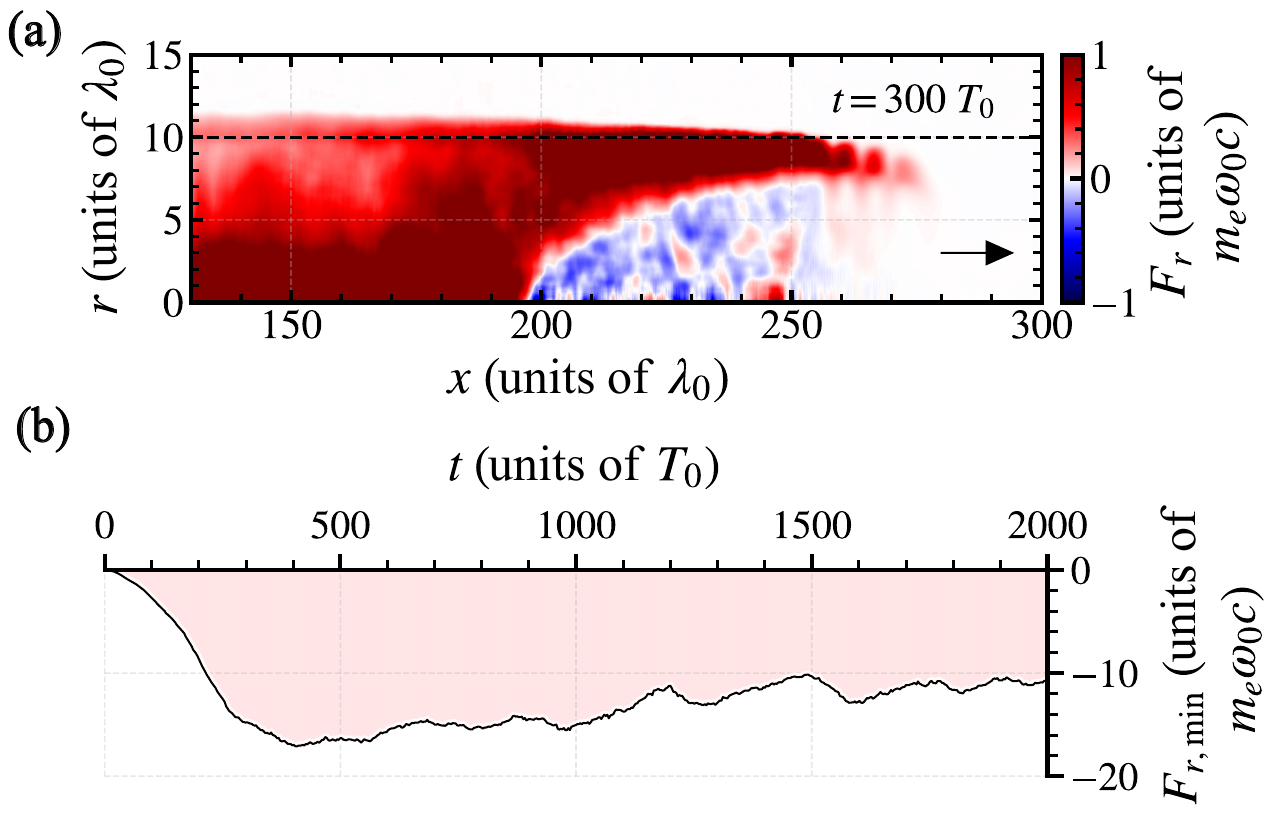}
    \caption{(Color online). (a) Channel fields contribution to the radial Lorentz force acting on positrons, $ F_r $, at the instant of time $ t = 300 \, T_0 $ obtained from the PIC simulation. The black dashed line and the arrow mark the initial channel radius, $ R $, and the direction of the laser pulse propagation, respectively. The colorbar is saturated. (b) Time evolution of the minimum of the Lorentz force displayed in panel (a), $ F_{r, \mathrm{min}} $. The values of $ F_{r, \mathrm{min}} $ are attained in the vicinity of the channel axis, which coincides with the electron density distribution inside the channel.}
    \label{fig:9}
\end{figure}

As closer elucidated in Ref.~\citenum{martinez2023}, a charge of accelerated electrons built up along the channel axis induces a negative charge separation field, which can overcome positive, self-generated radial electric field formed by the initial expulsion of electrons. This results in a formation of regions of space with focusing fields for positively charged particles, such as positrons, that co-propagate with driving laser over extended distances. In panel (a) of Fig.~\ref{fig:9} we plot the radial component of the Lorentz force acting on positrons, $ F_r = e \left( E_r - c B_{\phi} \right) $, induced by the self-generated radial electric, $ E_r $, and azimuthal magnetic, $ B_{\phi} $, fields in plasma at the time $ t = 300 \, T_0 $. The regions in which $ F_r $ is negative (displayed by the blue color) are focusing for positrons, enabling their guiding \cite{martinez2023}. The formation of such a guiding structure can be clearly seen in the simulation. 

In panel (b) of Fig.~\ref{fig:9} we further determine the temporal evolution of the minimum of $ F_r $, i.e., the magnitude of the force that is focusing for positrons. The value of minimum decreases after the entrance of the laser pulse to the channel and drops down to a global minimum ($ -17 \, m_e \omega_0 c $) at $ t \approx 400 \, T_0 $ (approximately at the time when the average electron density inside the channel is highest). After that, even though it shows variations over time, with alternating increases and decreases, the overall trend suggests a positive slope on average (to $ -10 \, m_e \omega_0 c $). Despite the decreasing magnitude of the positron focusing force, the guiding structure remains stable.

\section{\label{sec:conclusion} Conclusion}

In this work, we investigate the mechanism of DLA in preformed narrow cylindrical plasma channels using analytical methods and full-scale PIC simulation in quasi-3D geometry. The simulation demonstrates the generation of a high-charge electron source via DLA using a $ \mathrm{kJ} $-class driving laser as well as the formation of the guiding structure for positrons. We further focus on determining the parameters of this guiding structure and the locations at which the accelerated electrons originate. Considering the setup of a comparable channel diameter and the transverse size of the driving laser pulse, it turns out that the accelerated electrons are predominantly injected from the regions adjacent to the interface between the channel interior and the channel wall. 

Therefore, we formulate an analytical model based on the electron hydrodynamics that illustrates the release of electrons from the channel wall when irradiated by laser, the subsequent electron dynamics, and the corresponding evolution of the channel density profile in both non-relativistic and relativistic limits. In addition, the model allows to quantitatively predict the number of electrons released from the channel wall as well as the corresponding average electron number density inside the channel right after the passage of the laser pulse. These predictions are validated by comparison with the simulation results. 

We note that the situation described in this paper may change when using driving laser pulses of extreme intensity ($ > 10^{23} \ \mathrm{W / cm^2}$), i.e., when the effects of radiation-reaction force cannot be neglected and the laser-plasma interaction shifts towards the near-QED regime. In this case, the radiation-reaction trapping \cite{ji2014, vranic2018} of electrons begins to play an important role, so that the trapping efficiency of electrons with the origin near the channel axis (where the laser intensity is highest) may significantly increase.

In conclusion, the sources of high-energy and high-charge electrons generated by DLA reveal a strong potential for fundamental research as well as for various practical applications in diverse fields, including the production of bright gamma-ray and high-flux neutron sources \cite{rosmej2019, gunther2022}, exciting nuclear isomers \cite{feng2023}, seeding of the QED cascades \cite{jirka2017}, and the guiding of positrons \cite{martinez2023, maslarova2023}.

\acknowledgments   
 
P.V. would like to acknowledge the hospitality of the University of Lisbon. This work was supported by the NSF and Czech Science Foundation (NSF-GACR collaborative Grant No.~2206059 and Czech Science Foundation Grant No.~22-42963L). This work was supported by the Ministry of Education, Youth and Sports of the Czech Republic through the e-INFRA CZ (ID:90254). This work was supported by FCT grants CEECIND/01906/2018, PTDC/FIS-PLA/3800/2021 and FCT UI/BD/151560/2021. We acknowledge EuroHPC for awarding access to LUMI supercomputer.

\bibliographystyle{apsrev4-2}

\end{document}